\documentclass[fleqn,twoside]{article}
\usepackage{espcrc2}

\title{Strongly coupled N=1 SYM theory on the lattice}

\author{F.~Berruto and M.~Schwetz\\
[0.2cm]
Boston University - Department of Physics, 590 
Commonwealth Avenue, Boston MA 02215.}
       
\begin{document}

\begin{abstract}
We propose a strong coupling expansion as a possible tool to obtain
qualitative and quantitative informations about N=1 SYM theory. We
point out the existence of a mapping between strongly coupled lattice 
N=1 SYM theory and a generalized SO(4) antiferromagnetic spin system.
\vspace{1pc}
\end{abstract}

\maketitle
Though far from the scaling regime, the strong coupling limit of lattice
gauge theory exhibits confinement and chiral symmetry breaking 
and has often been used as a computational
scheme for understanding qualitative features of continuum theories 
such as QCD. Strongly coupled lattice gauge theories are intimately
related to quantum spin systems~\cite{qss,smit}. 
In Ref.~\cite{smit} Smit showed that 
the strong coupling effective hamiltonian of QCD regularized with Wilson 
fermions is a generalized $U(4N_f)$ 
antiferromagnet. 

N=1 supersymmetric Yang-Mills ($SYM$) theory is characterized by the
non-perturbative phenomena of $QCD$, such as color confinement and chiral
symmetry breaking. The main motivation of our analysis is to understand if
the low energy degrees of freedom of strongly coupled N=1 $SYM$ theory can be
described by some sort of antiferromagnetic spin model.
 
The seminal work by Seiberg~\cite{S} produced an avalanche of 
investigations of supersymmetric theories in a hope of unravelling 
the mysteries of quark confinement. 
Further twist in developments came from the work of Maldacena~\cite{Mal}
who managed to connect strongly coupled SYM theories with weakly
coupled supergravity. 
At the same time efforts have been made to give some ``experimental'' 
confirmations to the exciting theoretical
advances ( for models see e.g.~\cite{EHS}). 

However, well known difficulties prevent a definition of supersymmetric
theories on the lattice, such as the lack of lattice generators of the
Poincar\'e group and the unbalance between fermionic and bosonic degrees of
freedom (fermion doubling). The Wilson term cures this problem, but it
generates a bare mass term for the gluinos, which breaks explicitely 
both chiral symmetry and supersymmetry ($SUSY$). 
Curci and Veneziano~\cite{cv} proposed to
regularize on the lattice $SYM$ theories using Wilson fermions and claimed that
$SUSY$, like chiral symmetry, should only be recovered in the
continuum limit. For $SYM$ theory it is particularly fortunate that the
chiral limit and the $SUSY$ limit are one and the same. 
Consequently, many
$MC$ simulations of N=1 $SYM$ theory have been performed in the last few
years using Wilson fermions (for a recent review see e.g. \cite{Mon}). 
A pioneering study of N=1 $SYM$ theory in the strong coupling and large
$N_c$ limit was performed in~\cite{aga} using the Euclidean formalism and the
hopping parameter expansion in terms of random walks. Since the strong
coupling regime is far from the continuum limit, it is not guaranteed that
the strongly coupled lattice gauge theory, when extrapolated to the continuum,
would recover $SUSY$. However, the hope is that the strongly coupled lattice
theory could capture the gross features of the continuum strongly coupled 
$SYM$ theory. 
Therefore, we propose a strong coupling approach to N=1 $SYM$ theory in the 
hamiltonian formalism of lattice gauge theory using Wilson fermions and we
show that the theory is effectively described by an $SO(4)$ antiferromagnet. 

The N=1 $SU(N_c)$ $SYM$ lagrangian is 
\begin{equation}
\label{SYM}
{\cal L}=\frac{1}{2}\overline{\lambda}^a\gamma_{\mu}D^{ab}_{\mu}\lambda^b
+\frac{1}{4g^2}F_{\mu \nu}^a F_{\mu \nu}^a + 
\frac{\theta}{32\pi^2}F_{\mu \nu}^a\tilde{F}_{\mu \nu}^a\ 
\end{equation}
where the fermions obey the Majorana condition 
$\lambda=\gamma_0\overline{\lambda}^{\mbox{t}}$. 
The axial global symmetry $U(1)_A$ is broken by instantons to discrete 
$Z_{2N_c}$ symmetry, which in turn is broken to $Z_2$ symmetry by a gaugino
condensate. Correspodingly, there will be $N_c$ vacua which are
distinguished by the phase of the condensate 
$
\langle \overline{\lambda}\lambda \rangle=C\Lambda^3e^{\frac{2\pi i
k}{N_c}},\quad k=0, \ldots, N_c-1 .
$
One can add a gaugino mass term $m_g\overline{\lambda^a}\lambda^a$
to the SYM lagrangian~(\ref{SYM}) turning the model
into a {\it softly broken} SYM theory~\cite{EHS-soft}. 
This is the theory which lattice practitioners simulate.

In the Wilson formulation, the lattice action corresponding to Eq.~(\ref{SYM}) 
reads
\begin{equation}
S=\frac{2N_c}{g^2}\sum_P\left({\bf
    1}-\frac{1}{N_c}\mbox{Tr}\mbox{Re}U_P\right)
+S_F
\label{wga}
\end{equation}
where the gluino action is 
\begin{eqnarray}
S_F&=&\sum_{x,\mu}\mbox{Tr}[\overline{\lambda}(x)\frac{\left(\gamma_{\mu}-\hat{r}\right)}{2}U_{\mu}(x)\lambda(x+\hat{\mu})U_{\mu}^{\dagger}(x)\nonumber\\
& &-\mbox{h.c.}]+\sum_x \mbox{Tr}\overline{\lambda}(x)\left(m_g+4\hat{r}\right)\lambda(x)
\label{fa}
\end{eqnarray} 
and the trace $\mbox{Tr}$ is only over color. In $QCD$ regularized on the
lattice with Wilson fermions it is possible to introduce the $\theta$-vacua
dependence in the lattice action via the $\hat{r}$ Wilson
term~\cite{tv}. Since the action~(\ref{fa}) must be periodic in the $\theta$
angle with period $2\pi$, if one defines 
$\hat{r}=re^{i\theta\gamma_5/N_c}$ it is manifest that under a 
$U(1)_A$ rotation of the gluino field only the $Z_{2N_c}$ symmetry is
preserved.         

The hamiltonian of the theory ( rescaled by $g^2/2$ ) can be written as
\begin{equation}
H=H_E+\epsilon^2H_B+\epsilon(H_h+H_m)
\label{lh}
\end{equation}
where $\epsilon=1/g^2$ is the strong coupling expansion parameter, 
$g\rightarrow \infty$. 
In Eq.(\ref{lh}) $H_E=\sum_{x}E_{x,k}^aE_{x,k}^a$ is the chromo-electric
hamiltonian where $k=1,2,3$, $a=1,\ldots,N_c^2-1$ and 
$H_B=\sum_P \mbox{Tr}\mbox{Re}U_P$ is the chromo-magnetic hamiltonian. $H_B$ 
is of order $\epsilon^2$ 
and we shall neglect its contribution in our expansion, 
since it would contribute only at the fourth order. Therefore 
$H_E$ is the unperturbed hamiltonian that selects the vacuum of the theory 
and $H_h+H_m$ is the perturbation. 
The hopping hamiltonian $H_h=i\sum_x\left[H_h^F(x)-H_h^B(x)\right]$ 
in Eq.~(\ref{lh}) is written in terms of 
the forward and backward hopping hamiltonians $H_h^F=(H_h^B)^{\dagger}$
\begin{equation}
H_h^F=\sum_{x}\mbox{Tr}[
\lambda^{\mbox{t}}(x+\hat{k})(\gamma_0\hat{r}+
\alpha_k)U_k(x)^{\dagger}\lambda(x)U_k(x)]\ 
\label{hoppf}
\end{equation}
and 
$
H_{m_g}=2\sum_x \mbox{Tr}\left[\lambda^{\mbox{t}}(x)
\gamma_0\left(m_g+4\hat{r}\right)\lambda(x)\right]
$ 
is the gluino mass hamiltonian.

The ground state of $H_E$, $|0\rangle$, 
is defined by $E^a_k=0$ for every link, 
$H_E|0\rangle=0$, and is highly 
degenerate, since by applying an arbitrary number of times meson-like or 
baryon-like operators
\begin{equation}
M^{\Gamma}(x)=\frac{1}{2}\left[\lambda_{\alpha}^{a\ \mbox{t}}(x),
\lambda_{\beta}^a(x)\right]\
\Gamma^{\alpha\beta}\label{mls}
\end{equation}
\begin{equation}
B_{\alpha_1\ldots \alpha_{N_c^2-1}}(x)=\epsilon_{a_1\ldots a_{N_c^2-1}}\
\lambda_{\alpha_1}^{a_1}(x)\ldots 
\lambda_{\alpha_{N_c^2-1}}^{a_{N_c^2-1}}(x)\ 
\label{bls}
\end{equation}  
one does not change the energy of the eigenstates of $H_E$. 
In Eq.(\ref{mls}) $\Gamma$ is a matrix of the Dirac
Clifford algebra and in Eqs.(\ref{mls},\ref{bls}) the greek indices are Dirac
indices. The lowest energy state of $H_E$ has to be a color
singlet. 
The empty vacuum $|e.v.\rangle$ is an eigenstate of $H_E$ without gluinos 
$\lambda^a_x|e.v.\rangle=0$.    
At a given site a color singlet may be formed either by leaving the site
unoccupied or by creating on it a baryon-like color singlets by applying the
operator $B^{\dagger}_{\alpha_1\ldots \alpha_{N_c^2-1}}$~(\ref{bls}).
The fermion density operator (Eq.(\ref{mls}) for $\Gamma={\bf 1}$) is
\begin{equation}
\rho(x)=\left[\lambda_{\alpha}^{a\ \mbox{t}}(x),\lambda_{\alpha}^a(x)\right]/2
=n(x)-2(N_c^2-1)\ 
\label{fdo}
\end{equation}
where $n(x)=\lambda_{\alpha}^{a\ \mbox{t}}(x)\lambda_{\alpha}^a(x)$. 
The singlet creation operator (\ref{bls}) is symmetric in the Dirac indices
and therefore it carries an irreducible representation of $SU(N_c)$ with a
Young tableau with $N_c^2-1$ columns and one row. Since Fermi statistics 
does not allow two fermions with the same quantum numbers on the same 
site, we can at most have four singlets on a given site. Therefore, the allowed
representations of the $SO(4)$ algebra on one site are the empty singlet and
those with Young tableaux of $N_c^2-1$ columns and 1,2,3,4 rows respectively,
which are distinguished by the fermion numbers $\rho(x)=(N_c^2-1)(\nu-2)$, 
$\nu=0,1,2,3,4$. In order to reproduce the spectrum of a relativistic theory,
we constraint the fermion states to be half filled, i.e. for a ground
state of $H_E$, $\sum_x\rho(x)=0$. 
The states are labelled by the fermion density 
$\rho(x)$ and by the vector in the corresponding $SO(4)$ representation at
each site and are degenerate.        

At the first order in the strong coupling expansion the only non zero
contribution is given by the gluino mass
$E_0^{(1)}=\langle 0|H_{m_g}|0 \rangle$, since $\langle0|H_h|0\rangle=0$. 
The ground state degeneracy is removed at the second order in the strong 
coupling expansion
\begin{equation}
E_0^{(2)}=-2\langle 0|H_h^F\frac{1}{H_E}H_h^B|0\rangle
\label{so}
\end{equation}
Using the commutation relation $\left[H_E,H_h\right]=N_cH_h$ and integrating 
over the link variables~\cite{creutz}, one can write 
$E_0^{(2)}=J\sum_{x,k}\lambda^{a\ \mbox{t}}_{\alpha}(x+\hat{k})
\Sigma_{\alpha\beta}^k \lambda^b_{\beta}(x) \lambda^{b\ \mbox{t}}_{\gamma}(x)
\Sigma_{\gamma\delta}^{k\ \dagger}\lambda^a_{\delta}(x+\hat{k})$, 
where $J=1/2N_c(N_c^2-1)$ and $\Sigma^k=\gamma_0\hat{r}+\alpha_k$. 
Let us first consider the limit $r=0$. 
In this limit the spectrum of the lattice naive Majorana fermions 
exhibits doublers which explicitly break ``supersymmetry'', since the lattice
theory possess eight times more fermion states than boson states. The
doublers cancel the triangle anomaly, that on the contrary is manifest at
finite $r$ through explicit chiral symmetry breaking~\cite{smit}. 
Therefore, the axial 
$U(1)_A$ symmetry is preserved and instead of $N_c$ we expect only one
vacuum. However, 
this limit is particular instructive because the structure of the theory
becomes much simpler and we believe that one can obtain useful informations
to extrapolate to the $r\neq 0$ case. For $r=0$ it is particularly convenient
the spin diagonalization 
of the Majorana fermion fields 
$\lambda(x)=(\alpha_1)^{x_1}(\alpha_2)^{x_2}(\alpha_3)^{x_3}(\alpha_4)^{x_4}
\underline{\lambda}(x)$. 
In term of the transformed fields $\underline{\lambda}$, the $\Sigma^k$
matrices reduce to sign factors and therefore the
effective hamiltonian up to order $\epsilon^2$ becomes the spin hamiltonian
\begin{equation}
H_{\mbox{eff}}=H_{m_g}+H_J
\label{heff}
\end{equation}
where $H_{m_g}=2m_g\sum_x\epsilon(x)\underline{\lambda}^{a\ \mbox{t}}(x)
\gamma_0\underline{\lambda}^a(x)$ and
\begin{equation}
H_J=J\sum_{x,k}\left[
\rho(x)\rho(x+\hat{k})+S_{\alpha\beta}(x)S_{\beta\alpha}(x+\hat{k})\right]
\label{hj}
\end{equation}
In Eq.~(\ref{hj}) $\epsilon(x)=(-1)^{x_1+x_2+x_3}$, we dropped a constant term and 
we introduced the $SO(4)$ spin operators ($\alpha\neq \beta$) 
$
S_{\alpha\beta}(x)=\frac{1}{2}\left[
\underline{\lambda}^{a\ \mbox{t}}_{\ \alpha}(x),
\underline{\lambda}_{\ \beta}^a(x) \right]
$.
The effective hamiltonian~(\ref{heff}) conserves the gluino number
locally, $\left[\rho(x),H_{\mbox{eff}}\right]=0$, while at higher orders 
in the
strong coupling expansion $H_{\mbox{eff}}$ contains also the 
baryon-like field (\ref{bls}) with nearest neighbor couplings such that the
gluino number is only conserved globally. 
Therefore, $\rho(x)$ is a good quantum number to classify the states of the 
hamiltonian (\ref{heff}). The generators $S_{\alpha \beta}(x)$ 
commute with $\rho(x)$ and the subspace with given
$\rho(x)$ transforms according to an irreducible representation of $SO(4)$. 
The ground states of the hamiltonian (\ref{heff}) are generalized N\'eel
states, where the spin points in one direction 
(e.g. the $\gamma_0$ direction) on the even sites and in
the opposite direction on the odd sites. The condensate $\langle 
\overline{\lambda}\lambda\rangle=1/V \sum_x
\epsilon(x)\langle M^{\gamma_0}(x)\rangle$ is non-zero on a N\'eel state and 
chiral symetry is spontaneously broken. 
The $r\neq 0$ case is more complicated, however with a Fierz
rearrangement~\cite{smit} one can express the effective hamiltonian 
in terms of meson-like operators~(\ref{mls}) 
and again one obtains a generalized $SO(4)$ 
antiferromagnet which will now have more
than one ground state depending on the magnitude of $m_g$ and $r$. Perturbing 
around the $r$=0 antiferromagnetic vacuum, one could compute the 
condensate and the mass spectrum.

\end{document}